\documentclass[10pt,twocolumn,letterpaper]{article}

\usepackage{cvpr}              

\usepackage{graphicx}
\usepackage{amsmath}
\usepackage{amssymb}
\usepackage{booktabs}

\usepackage[pagebackref,breaklinks,colorlinks]{hyperref}

\usepackage[capitalize]{cleveref}
\crefname{section}{Sec.}{Secs.}
\Crefname{section}{Section}{Sections}
\Crefname{table}{Table}{Tables}
\crefname{table}{Tab.}{Tabs.}

\def\cvprPaperID{*****} 
\def\confName{CVPR}
\def\confYear{2022}

\begin{document}

\title{Modality Dropout for Multimodal Device Directed Speech Detection using Verbal and Non-Verbal Features}

\author{Gautam Krishna
\quad
Sameer Dharur
\quad Oggi Rudovic 
\quad Pranay Dighe \\
\quad Saurabh Adya 
\quad Ahmed Hussen Abdelaziz
\quad Ahmed H Tewfik \\
Apple
}


\maketitle

\begin{abstract}
Device-directed speech detection (DDSD) is the binary classification task of distinguishing between queries directed at a voice assistant versus side conversation or background speech. State-of-the-art DDSD systems use verbal cues, e.g acoustic, text and/or automatic speech recognition system (ASR) features, to classify speech as device-directed or otherwise, and often have to contend with one or more of these modalities being unavailable when deployed in real-world settings. In this paper, we investigate fusion schemes for DDSD systems that can be made more robust to missing modalities. Concurrently, we study the use of non-verbal cues, specifically prosody features, in addition to verbal cues for DDSD. We present different approaches to combine scores and embeddings from prosody with the corresponding verbal cues, finding that prosody improves DDSD performance by upto \textbf{8.5\%} in terms of false acceptance rate (FA) at a given fixed operating point via non-linear intermediate fusion, while our use of modality dropout techniques improves the performance of these models by \textbf{7.4\%} in terms of FA when evaluated with missing modalities during inference time.
\end{abstract}


\section{Introduction}
\label{sec:intro}

Voice assistants (VAs) have recently become ubiquitous in controlling smart devices and in information retrieval. VAs are used in challenging acoustic environments, where noise levels are high. In such environments, wake word spotting and ASR systems sometimes misrecognize noise as legitimate words and hallucinated queries, respectively. To improve the user experience, it is crucial to design an automatic system that distinguishes between device-directed queries and side conversations or ambient noise.  

We refer to this binary classification task that detects device-directed queries and rejects unintended speech as Device Directed Speech Detection (DDSD). 

Non-verbal cues are essential for a successful human-human interaction \cite{psychology}. We typically use acoustic and visual non-verbal signals to communicate additional information that is not contained in spoken words. For example, prosody features are used to improve the mechanics of interaction between humans by regulating turn-taking and highlighting the nature of sentences, e.g., changes in pitch could indicate whether a sentence is a statement or a question. DDSD systems that have been studied in literature \cite{kumar2020building,dighe2020lattice,garg2021streaming,mallidi2018device, RudovicTCN} rely solely on verbal signals, such as text or speech. Furthermore, prior work pre-supposes the availability of all multi-modal data at inference time, which often occurs in real-world settings. In this paper we attempt to bridge both these gaps by building DDSD systems that include prosody features intuitive to a conversation and also train these fusion models to be robust to missing modalities at inference time, which is a critical requirement for systems deployed in complex acoustic environments. 

The prosody features used in this work are pitch, voicing, jitter, shimmer, and voice activity detection (VAD). The prosody-based device directedness classifier is a Gated Re-\\current Unit (GRU)-based neural network \cite{chung2014empirical} that consumes a sequence of the concatenated prosody features and makes a query-level directedness decision. The GRU summarizes the temporal prosody features that are extracted from the entire query. We compare the prosody-based DDSD model to DDSD models that use conventional verbal cues, such as acoustic, text, and ASR.

We refer to the output posteriors and the penultimate layer embeddings of DDSD models as the directedness features. We investigate the impact of combining prosody directedness features with conventional verbal signal-based models on the accuracy of DDSD. We examine different fusion schemes, such as linear and non-linear late fusion as well as intermediate fusion, and find that using such non-verbal directedness significantly improves DDSD performance.

Our main contributions are: 1) We introduce novel DDSD models that augment traditional verbal features with non-verbal prosody. To the best of our knowledge, this is the first such work to prove the efficacy of non-verbal features in the context of DDSD. 2) Using different fusion schemes, we improve the accuracy of DDSD when using prosody directedness features relative to the baseline DDSD that uses verbal cues only. 3) We show that our fusion schemes can be made more robust to missing data by implementing modality dropout in fusion models, where we stochastically drop features during training time based on a predefined probability value. 

The rest of the paper is organized as follows: In Sec.\ref{sec:RW}, we discuss prior work related to our problems. Sec.\ref{sec:DDNF} describes the prosody features and prosody feature-based DDSD that we use in this study. In Sec.\ref{sec:FS}, we outline the different verbal and non verbal fusion schemes.
The experiments used to evaluate the model performance and the results are described in Sec.\ref{sec:Exp}. Finally, we conclude the paper and give an outline of future work in Sec.\ref{sec:conc}.

\section{Related Work}
\label{sec:RW}

\subsection{DDSD with Verbal Features}
The problem of device-directed speech detection, and variants thereof \cite{https://doi.org/10.48550/arxiv.2008.00508}, has been extensively studied in recent work – principally using verbal features from speech and text \cite{kumar2020building}, and combinations of the two with various representations and supervision techniques \cite{dighe2020lattice, mallidi2018device, RudovicTCN, Ladhak2016}. In this work, we augment this verbal feature set with non-verbal features appropriate to this task, i.e: prosody signals. Our work differs from \cite{shriberg2012learning} in our choice of prosodic modeling features, fusion schemes and modality dropout techniques aimed at robustness. Compared to the acoustic-based DDSD model proposed in \cite{RudovicTCN}, where log-mel filterbank features are used to train a spoken language understanding (SLU) model for the device-directedness task, the prosody-based DDSD model proposed in this study exploits the fact that humans use different speech style when communicating with human versus when communicating with a VA. For example, humans tend to talk louder, slower, over-articulate, and  avoid having stops or speech fillers, such as `Uh' and `um', when they issue a command to a VA. In other words, while the acoustic-based DDSD focuses on what was said, prosody-based DDSD focuses on how an query was said to predict device directedness. As shown in Section \ref{sec:results}, combining both acoustic-based and prosody-based DDSD improves the perfor\\mance of DDSD. 

In \cite{talking-faces} authors apply modality dropout techniques to performance-driven facial animation using visual and acoustic features, while our work focuses on DDSD using prosody, text and acoustic features.

\section{DDSD using Non-verbal Features}
\label{sec:DDNF}
\subsection{Prosody Features}

We use off-the-shelf algorithms to extract five-dimensional prosody feature vectors that contain – pitch \cite{ghahremani2014pitch}, voicing \cite{ghahremani2014pitch}, jitter \cite{farrus2007jitter}, shimmer \cite{farrus2007jitter}, and voice activity detection (VAD) scores at a sampling frequency of 100 Hz. The internal VAD model that we used to get VAD scores consists of a binary speech/silence classifier, followed by a two state (speech/silence) Hidden Markov Model (HMM). The binary speech/silence classifier is a convolutional neural network based model (CNN) followed by 4 fully-connected layers leading to the final 2-output softmax scores. Then the 2 outputs from the speech/silence classifier are treated as observation probabilities for a 2-state HMM.  

\subsection{Prosody-based DDSD}
We use a single layer of Gated Recurrent Units (GRUs) \cite{chung2014empirical} applied to the temporal prosody features. The input features are pre-processed by a masking layer that enables the GRU layer to skip the processing of the padded zeros. The GRU layer contains 128 hidden units. The last time-step output of the GRU layer normalized, followed by a dropout regularization \cite{srivastava2014dropout}, with dropout rate of 0.2. Finally a dense layer with a single hidden unit and a sigmoid activation function is used to estimate a query-level device directedness probability, i.e, the probability whether a query is directed towards a device or not. 

The model is trained for 50 epochs using Adam optimizer \cite{kingma2014adam} with a learning rate of 0.001. We further use gradient clipping \cite{zhang2019gradient} to reduce over-fitting. 

The batch size is set to 150 and a weighted binary cross-entropy loss function is used to compensate for the class imbalance in the training data. The models are trained using four GPUs. The model contains 50K learnable parameters. 

\section{Fusion Schemes}
\label{sec:FS}

The multi-modal fusion realm continues to attract substantial research interest in different machine learning applications \cite{yang2023overview}. Fusion models are usually categorized into three categories: early or feature fusion, late or score fusion, and intermediate or embedding fusion schemes \cite{boulahia2021early}.  We focus on the late and intermediate fusion schemes, as the single-modality models used here are highly optimized and trained with modality specific corpora that do not necessarily contain all other modalities. In particular, we aim at combining our non-verbal directedness feature, i.e., prosody, with three verbal features, which are acoustic, text, and ASR directedness features.  

As an acoustic-based DDSD approach, we adopt the recently proposed streaming Acoustic False Trigger Mitigator (aFTM) ~\cite{RudovicTCN}, designed for binary classification of directed speech based on acoustic data only. The acoustic-based DDSD model is trained using 40-dimensional filter-bank features.
 As an ASR-based DDSD approach, we use the modified form of LatticeRNN \cite{jeon2019latrnn}. As a text-based DDSD approach, we use a DNN-based classifier head on top of a fine-tuned BERT-like model \cite{tenney2019bert}.   

In the case of late fusion, we investigate applying linear and non-linear functions/layers to the output posteriors/scores of the different verbal and non-verbal DDSD models. In the intermediate fusion, we apply non-linear layers to the embeddings that are extracted from the penultimate layers of the corresponding single-modality DDSD models. The details of the late and intermediate fusion schemes are described further in Sections \ref{sec:lateFus} and \ref{sec:intFus}. The modality dropout concept is explained in \ref{sec:drop}. 

\subsection{Late Fusion}
\label{sec:lateFus}

We denote the scores of the acoustic-, text-, ASR-, and prosody-based DDSD models as $S_\text{a}$, $S_\text{t}$, $S_\text{asr}$, and $S_\text{p}$, respect-\\ively. For the late fusion scheme, we compare the simple linear averaging, denoted by AVG, and the non-linear score fusion using deep neural networks (DNNs), denoted by SL. 

Now we describe the architecture of the SL fusion model. The SL model accepts scores from individual DDSD models as inputs in parallel branches. Each score is pre-processed by an inverse softmax layer followed by a  fully connected dense layer with 128 hidden units and Tanh non-linearity. The inverse softmax layer is applied to the input scores to learn to recover the lost logits which contains richer information. The post-processed scores of all single modality DDSD models are concatenated and then passed through another dense layer with ReLU \cite{agarap2018deep} non-linearity containing 128 hidden units followed by layer normalization and a dense layer with a sigmoid activation function, which outputs the final combined score. 

To prove the value of the non-verbal features, we compare the linear and non-linear late fusion baselines that utilize the verbal scores only, i.e., $S_\text{a}$, $S_\text{t}$, and $S_\text{asr}$, with the corresponding ones that also incorporate non-verbal prosody scores.   

\subsection{Intermediate Fusion}
\label{sec:intFus}

Intermediate layers of single-modality DNN models usually contain richer information than just the output scores. Fusing the output of the intermediate layers is usually referred to as intermediate fusion \cite{boulahia2021early}. In this paper, we apply intermediate or embedding level (EL) fusion to the output of the penultimate layers of the single-modality DDSD models. The fusion network architecture is similar to late fusion network except that the inputs are embeddings and inverse-softmax layers are removed.  

The embedding from the prosody DDSD model is the 128-dimensional output vector of the last time step of the GRU layer, denoted by $E_\text{p}$. The acoustic- and text-based embeddings are of dimension 256 and 128 respectively, denoted by $E_\text{a}$ and $E_\text{t}$, respectively. The baseline for embe-\\dding fusion experiments utilizes only embeddings from the verbal DDSD models, i.e., $E_\text{a}$, $E_\text{t}$, and $E_\text{asr}$. $E_\text{asr}$ is of dimension 16. 

Similar to the late fusion experiments, we compare the baseline results with the intermediate fusion results that incorporate non-verbal embeddings, i.e. $E_\text{p}$. Both non-linear late and intermediate fusion models are trained for 50 epochs using the Adam optimizer with batch size 150 and weighted binary cross-entropy as the loss function. 

\subsection{Modality Dropout}
\label{sec:drop}

It is often the case that all the DDSD models may not produce outputs at the same time for all possible inputs, which requires the fusion models to be able to handle missing data upon deployment. We investigate an input modality dropout technique during training to improve the robustness to missing data for the late and intermediate fusion models. Based on hyper-parameter tuning experiments performed using our validation data split, we determine the dropout probability values to be used with acoustic-, Prosody-, ASR- and text-based DDSD model scores and embedding input branches/layers of both Late and Intermediate fusion models. The main idea here is to apply dropout regularization to the input layers of the fusion models. 

Our results demonstrate that adding input layer modality dropout make our fusion schemes more robust during inference time to missing modalities. We evaluate the modality dropout scheme on our standard eval set which contains no missing data, as well as on a variation of the eval data missing 30\% of scores and embeddings randomly dropped for each of the DDSD models. The dropped scores were replaced with a value of negative one and dropped embeddings were replaced with an array filled with -99999 of dimension corresponding to the DDSD embedding. 

\section{Experiments}
\label{sec:Exp}
\subsection{Dataset}
We collect a multi-modal dataset from demographically diverse English native speakers with consent, which is tailored to simulate a naturally occurring multi-party conversation towards solving a task together (e.g, cross-word puzzles). During the interaction, the participants ask the VA questions that they might not know the answers for. The participants and moderators use the answers to complete the task at hand. During the interaction, the VA-equipped devices record all multi-modal signals of the participants. 

The dataset consists of over 19,000 audio recordings of 530 hours from 1,300 participants. The conversations are clipped into roughly 245k segments and then annotated by human graders with information about the interaction including device-directedness. The dataset segments are split into training, validation and test splits in a roughly 60:20:20 ratio. The training and validation splits are further divided into component and fusion subsets. The component splits are used to train the single modality DDSD models while the fusion splits are used to train the fusion models (Table \ref{tab:data}). All five splits are independent with no-overlap in data or speakers between the splits, which is an important consideration to gauge the generalizability of our models.

\begin{table}[!ht]
  \caption{Number of queries in the training, validation, and test splits for developing the component and fusion models}
  \label{tab:data}
  \centering
  \resizebox{6cm}{!}{
  \begin{tabular}{ cccc}
    \toprule
    \multicolumn{1}{c}{\textbf{Labels}} & 
    \multicolumn{1}{c}{\textbf{Train Comp/Fus}} &
    \multicolumn{1}{c}{\textbf{Val Comp/Fus}} &
    \multicolumn{1}{c}{\textbf{Test}} \\
    \midrule
                           Directed  &  $5.2k/3.4k$ & $2.6k/1.5k$ & $3.1$k \\
                           Not Directed  &  $30k/18k$ & $12.5k/7.4k$ & $17k$ \\
    \bottomrule
  \end{tabular}
  }
\vspace{-4mm}
\end{table}

\subsection{Results}
\label{sec:results}
DDSD models are binary classifiers that accept queries as directed to the VA or reject them as un-directed side speech. In all experiments, we use the equal error rate (EER) and false acceptance rate (FA) at 10 \% false reject rate (FR) (FA@10\% FR) measured on the test set as the evaluation metrics. Lower FA @FR=10\% and EER values mean better DDSD performance. Table \ref{tab:single} shows the performance of single-modality DDSD models on the test set. The performance of the prosody-based DDSD model is better than a random classifier trained using gaussian noise. However, they are better combined with the verbal DDSD models, which have superior performance on this task. 

\begin{table}[!ht]
  \caption{EER and FA@10\% FR of single-modality DDSD models evaluated on the test set containing \textbf{no missing modalities}}
  \label{tab:single}
  \centering
  \resizebox{6cm}{!}{
  \begin{tabular}{ rcc }
    \toprule
    \textbf{Modality} & \textbf{FA@10\% FR [\%]} & \textbf{EER [\%]} \\
    \midrule
                            Text   & $22.2$ & $13.7$             \\
                            Acoustic    & $30.5$ & $17.8$             \\
                          ASR     & $20.5$ & $13.1$       \\
                         Prosody   & $33.6$ & $17.3$             \\
                         \midrule

                         Random Classifier   & $89.9$ & $50.6$  
                         \\
    \bottomrule
  \end{tabular}
  }
\vspace{-2mm}
\end{table}

We then identify the most optimal fusion schemes to combine DDSD signals and report, in Table \ref{tab:late}, that intermediate or embedding-level fusion gives better fusion results than late fusion of scores or a blind averaging of scores. This is intuitively expected since the embeddings contain richer information that the fusion neural networks can exploit compared to the output posteriors. 

\begin{table}[!ht]
  \caption{EER and FA@10\% FR of linear (AVG) and non-linear (SL) late fusion models evaluated on the test set containing \textbf{no missing modalities}}
  \label{tab:late}
  \centering
  \resizebox{7cm}{!}{
  \begin{tabular}{ rcc}
    \toprule
    \textbf{Fusion model} &  \textbf{FA@10\% FR[\%]} & \textbf{EER[\%]} \\
    \midrule
                            AVG($S_\text{a}$, $S_\text{t}$, $S_\text{asr}$)  & $14.5$ & $11.1$             \\

                         SL($S_\text{a}$, $S_\text{t}$, $S_\text{asr}$)  & $13.8$ & $11.0$    \\
                        \textbf{EL($E_\text{a}$, $E_\text{t}$, $E_\text{asr}$)}  & $\textbf{7.1}$ & $\textbf{9.2}$            \\

    \bottomrule
  \end{tabular}
   }
  \vspace{-2mm}
\end{table}

Additionally, the results in Table \ref{tab:intermediate} show that using prosody  signal improves the performance of DDSD compared to the baseline models that only use verbal signals by around 8.4\% and 3.2\% relative reduction in FA@10\%FR and EER, respectively as evaluated on the full test set. Furthermore, the final row of the table indicates that a model trained with modality dropout technique (MD) (as described in Section \ref{sec:drop}) provides an additional improvement in terms of reducing the false acceptance rate. 

\begin{table}[!ht]
  \caption{EER and FA@10\% FR of intermediate fusion models (EL) evaluated on the test set containing \textbf{no missing modalities}}
  \label{tab:intermediate}
  \centering
  \resizebox{7cm}{!}{%
  \begin{tabular}{ rcc}
    \toprule
    \textbf{Fusion model} &  \textbf{FA@10\% FR[\%]} & \textbf{EER[\%]} \\
    \midrule
                            EL($E_\text{a}$, $E_\text{t}$, $E_\text{asr}$)  & $7.1 $&$9.2$             \\  
                            EL($E_\text{a}$, $E_\text{t}$, $E_\text{asr}$, $E_\text{p}$)  & $6.5 $&$8.9 $       \\
                            \textbf{EL($E_\text{a}$, $E_\text{t}$, $E_\text{asr}$, $E_\text{p}$) + MD} & $\textbf{5.7}$ & $\textbf{8.9}$ \\
    \bottomrule
  \end{tabular}
  }
  \vspace{-2mm}
\end{table}

Finally, Table \ref{tab:drop} 
show the results for intermediate fusions with and without modality drop-out algorithm when evaluated on our test set containing missing data where we forcibly drop 30\% of scores and embeddings for each of the DDSD models at random and replaced them with dummy values as described before in \ref{sec:drop}. We see that the use of modality dropout technique makes the fusion models more robust to missing data problem during inference time. The results also demonstrate the value of prosody features in improving DDSD even in the presence of missing modalities at inference time.

\begin{table}[!ht]
  \caption{EER and FA@10\% FR of intermediate fusion models (EL) evaluated on the test set containing 30\% \textbf{missing modalities}}
  \label{tab:drop}
  \centering
  \resizebox{7cm}{!}{
  \begin{tabular}{ rcc}
    \toprule
    \textbf{Fusion model} &  \textbf{FA@10\% FR[\%]} & \textbf{EER[\%]} \\
    \midrule
                            EL($E_\text{a}$, $E_\text{t}$, $E_\text{asr}$)  & $21.94 $&$11.45 $       \\
                            EL($E_\text{a}$, $E_\text{t}$, $E_\text{asr}$, $E_\text{p}$)  & $11.46 $&$10.43 $       \\
                            \textbf{EL($E_\text{a}$, $E_\text{t}$, $E_\text{asr}$, $E_\text{p}$) + MD}    & $\textbf{10.61} $&$\textbf{10.17} $       \\
    \bottomrule
  \end{tabular}
  }
  \vspace{-2mm}
\end{table}

\section{Conclusions}
\label{sec:conc}
In this paper, we study device directedness speech detection (DDSD) model that use prosody, to distinguish between speech directed to a device and side speech or ambient noise. We have demonstrated that the performance of DDSD models that rely solely on conventional verbal signals, such as acoustic, text, and ASR features, can be improved by incorporating  prosody features using late and intermediate fusion schemes. Such improvements in DDSD models could enhance user privacy by reducing or eliminating the access of downstream components of voice assistant systems to users' private conversations, while also providing a more accessible and intuitive way of interacting with voice assistants. We further showed that the fusion schemes can be made more robust to missing data during inference time with the help of modality dropout (MD) during training and the technique also reduces false acceptance rate even in the absence of missing data while performing fusion. 


{\small
\bibliographystyle{ieee_fullname}
\bibliography{egbib}
}

\end{document}